\documentclass[lettersize,journal]{IEEEtran}
\usepackage{amsmath,amsfonts}
\usepackage{algorithm}
\usepackage{array}
\usepackage[caption=false,font=normalsize,labelfont=sf,textfont=sf]{subfig}
\usepackage{textcomp}
\usepackage{stfloats}
\usepackage{url}
\usepackage{verbatim}
\usepackage{graphicx}
\usepackage{filecontents}
\usepackage{graphicx}
\usepackage{pifont}
\usepackage{hyperref}
\usepackage{shortcuts}
\usepackage{color}
\usepackage{amsmath}
\usepackage{booktabs}
\usepackage{soul}
\usepackage{enumitem}
\usepackage{array}
\usepackage{tabularx} 
\usepackage{subcaption}
\usepackage{float}
\usepackage{amsmath}
\usepackage{amsfonts}
\usepackage{threeparttable}
\usepackage[T1]{fontenc}
\usepackage{framed} 
\usepackage{multirow}
\usepackage{makecell}
\usepackage{xcolor}
\usepackage{tikz}
\usepackage{diagbox}
\usepackage{algpseudocode}
\usepackage{mdframed}
\usepackage{framed}
\usepackage{wrapfig,lipsum}
\usepackage[strict]{changepage}
\usepackage{url}
\definecolor{heatlow}{rgb}{0.8,0.8,1}
\definecolor{heatmid}{rgb}{0.5,0.5,1}
\definecolor{heathigh}{rgb}{0,0,1}

\definecolor{formalshade}{RGB}{250,250,255}
\definecolor{LightSkyBlue}{RGB}{161,178,208}

\newenvironment{formal}{%
  \MakeFramed{\advance\hsize-\width\FrameRestore}%
  \noindent\hspace{-4.55pt}
  \begin{adjustwidth}{}{7pt}%
}
{%
  \end{adjustwidth}\endMakeFramed%
}
\begin{document}

\title{A Multi-Store Privacy Measurement of Virtual Reality App Ecosystem}

\author{Chuan Yan,~\IEEEmembership{Staff,~IEEE,}
\thanks{This paper was produced by the IEEE Publication Technology Group. They are in Piscataway, NJ.}
\thanks{Manuscript received April 19, 2021; revised August 16, 2021.}}
\makeatletter
\newcommand{\linebreakand}{%
  \end{@IEEEauthorhalign}
  \hfill\mbox{}\par
  \mbox{}\hfill\begin{@IEEEauthorhalign}
}
\makeatother

\author{\IEEEauthorblockN{Chuan Yan,}
\and
\IEEEauthorblockN{Zeng Li,}
\and
\IEEEauthorblockN{Kunlin Cai,}
\and
\IEEEauthorblockN{Liuhuo Wan,}
\and
\IEEEauthorblockN{Ruomai Ren,\\}
\and
\IEEEauthorblockN{Yiran Shen,}
\and
\IEEEauthorblockN{Guangdong Bai\textsuperscript{}}\thanks{
\noindent\rule{0.9\columnwidth}{0.4pt} \\
\noindent
\textit{
\textbullet \quad C. Yan, L. Wan, R. Ren and G. Bai are with The University of Queensland, Australia.\\
\textbullet \quad Z. Li and Y. Shen are with Shandong University.\\
\textbullet \quad K. Cai is with The University of California, Los Angeles (UCLA)\\
}}
}


\maketitle

\begin{abstract}
Virtual Reality (VR) has gained increasing traction among various domains in recent years, with major companies such as Meta, Pico and Microsoft launching their application stores to support third-party developers in releasing their applications~(or simply \emph{apps}).  
These apps offer rich functionality but inherently collect privacy-sensitive data, such as user biometrics, behaviors, and the surrounding environment. 
Nevertheless, there is still a lack of domain-specific regulations to govern the data handling of VR apps, resulting in significant variations in their privacy practices among app stores.

In this work, we present the \emph{first} comprehensive multi-store study of privacy practices in the current VR app ecosystem, covering a large-scale dataset involving \appnum apps collected from \storenum major app stores. 
We assess both \emph{declarative} and \emph{behavioral} privacy practices of VR apps, using a multi-faceted approach based on natural language processing, reverse engineering, and static analysis.
Our assessment reveals significant privacy compliance issues across all stores, underscoring the premature status of privacy protection in this rapidly growing ecosystem.
For instance, one third of apps fail to declare their use of sensitive data, and 21.5\% of apps neglect to provide valid privacy policies. 
Our work sheds light on the \emph{status quo} of privacy protection within the VR app ecosystem for the first time. 
Our findings should raise an alert to VR app developers and users, and would encourage store operators to implement stringent regulations on privacy compliance among VR apps.   

\end{abstract}

\section{Introduction}
Virtual Reality (VR) has exhibited remarkable growth in recent years, driven by the convergence of advancements in software and hardware technologies. 
It creates computer-generated digital environments, providing users with immersive experiences through specialized devices such as head-mounted displays. 
VR has gained widespread popularity across a broad spectrum of domains, such as gaming and entertainment, education and distance learning, remote collaboration, healthcare and rehabilitation. 
The global VR market is projected to achieve a compound annual growth rate of 14\% from 2023 to 2032, as highlighted by a recent report~\cite{globel2023virtual}. 

The growth of VR can be partly attributed to the continuously enriching content offered by flourishing VR apps. 
Inspired by the great success of the mobile app ecosystem, VR service providers, such as Meta Oculus and Pico~\cite{vr2025global}, have established app stores to facilitate third-party developers in releasing their apps to users. 
These stores have become a hub for users to access a wide variety of VR experiences. 
The growing popularity of VR apps also introduces significant privacy risks. 
Their immersive nature entails access to extensive privacy-sensitive data through multi-modal sensors deployed in VR devices, including user biometrics, movements, behaviors, and even the surrounding environment. 
This unprecedented level of access to sensitive data makes VR apps particularly vulnerable to privacy breaches.

A crucial line of research has focused on understanding privacy leakage related to VR-specific data.  
It reveals that the leakage of VR-specific data, such as virtual identities, social interactions, and biometrics~\cite{heruatmadja2023biometric}, can lead to severe privacy issues such as precise identification of users or detailed profiling of their habits~\cite{NairGSO23Exploring,fernandez2022life, buck2021privacy,happa2021privacy}.
Some recent studies~\cite{trimananda2022ovrseen,guo2024empirical} have also begun exploring the privacy practices of VR apps' behavior within individual stores. 
However, the broader landscape of \emph{how effectively VR app developers and app store operators can safeguard user privacy and rigorously enhance their measures} remains largely unexplored.

Evaluating privacy risks in the VR domain is a challenging task. 
At the app level, existing compliance checkers~\cite{zhang2024navigating,zimmeck2016automated} 
are not directly applicable due to their limitation in integrating VR domain-specific knowledge. 
Although regulations like the European Union (EU) General Data Protection Regulation (GDPR)~\cite{gdpr} and California Consumer Privacy Act (CCPA)~\cite{ccpa} have been proposed to impose obligations on data controllers and processors, they only provide high-level guidelines. 
Effective compliance checking in the VR domain requires a combination of these regulations with domain-specific insights to understand the real practices~\cite{yan2024on}. 
Moreover, VR apps encompass more complex use cases and designs, equipped with additional sensors to collect  VR-specific sensitive data~\cite{NairGSO23Exploring}, such as eye tracking and hand movement data, which are beyond the scope of existing analyzers for traditional mobile device apps~\cite{samarin2023lessons,fan2020empirical}. 
For instance, the implications of motion data leaks in VR can vary significantly~\cite{su2024remote,nair2023truth}, and the collection of additional child data in these environments may lead to severe consequences~\cite{spiegel2018ethics,plechata2019age,maloney2020anonymity}. 
The complexity of the VR ecosystem also poses significant challenges for large-scale analysis. 
VR apps can be built using various APIs in different game engines, making them difficult to analyze. These apps are released through different app stores, each targeting different users, offering different functionality types, and adhering to their own store-specific policies that explain VR-specific requirements, such as Meta Quest's Virtual Reality Check.

\paragraph{Our Work} In this work, we conduct the first comprehensive study on the privacy practices of VR apps in the current app ecosystem. 
By combining domain-specific knowledge collected from VR research and various standards in VR privacy, we summarize two categories of VR-specific privacy-sensitive data and five VR-domain privacy concerns. 
Using this domain-specific knowledge, we create a compliance-checking pipeline and conduct a multi-store analysis to understand the privacy practices of existing VR apps in the wild. 
This approach addresses existing challenges and provides an effective and comprehensive understanding of VR privacy compliance issues.



Our multi-store analysis targets \storenum major VR stores, including Oculus, Viveport, Pico, Microsoft, and PlayStation,
which represent over 80\% of the global market share of VR devices as of 2025~\cite{vr2025global}.
This allows us to identify the different privacy problems caused by various store designs and highlight the shared problems that the community should pay more attention to. 
Moreover, the results of the analysis will enable us to perform adaptive compliance checking, which can be used in the majority of future VR apps.


Our analysis includes investigations of the apps' privacy policy documents and privacy meta-information disclosed through different app stores.
The former serves as the formal disclosure and contractual agreement between the user and the developer~\cite{lim2022mine,buck2021privacy,huang2021systematic,lopez2016method}, and the latter serves as the portal for the user to glean insights into the app's data collection practices. 
We aim to explore two research questions~(RQs) regarding these declarative documents. 
\emph{\textbf{RQ1)} What is the status of VR apps in providing users with adequate documents to declare their data handling practices?} 
This RQ investigates whether comprehensive documents are available and easy-to-interpret by average users, with a horizontal comparison of the types of meta-information mandated by different VR app stores~(\textbf{Section~\ref{pp_analysis}}). 
This comparison addresses the gap in current research that analysis of a single VR app store~\cite{guo2024empirical} is not sufficient and allows for a comprehensive understanding of the real VR store ecosystem, which contains multiple stores.
\emph{\textbf{RQ2)} Do existing VR apps provide complete and accurate content in their declarative documents?} 
This RQ assesses whether the contents provided cover the key components of concern to users and legislators, such as permission requests, data collection, and data retention. 
This assessment combines high-level regulations with VR domain-specific knowledge to provide an in-depth analysis of the thoroughness of privacy content in the VR domain, which is crucial for ensuring user trust and compliance with legal standards.
We also specifically check child user privacy protections, considering the large child user base of VR apps~(\textbf{Section~\ref{sec:doc_content}}). 

For privacy-relevant behaviors,
we aim to explore \emph{\textbf{RQ3)} what is the status of existing Android VR apps' actual behaviors in complying with their privacy policies?}
To this end, we construct the \emph{behavioral profiles} that represent apps' access to VR-specific data. 
We develop an analysis framework specifically designed for Android VR apps, given their wide use and the availability of their executable files~(i.e., \emph{.apk} files), which represents the mainstream of the VR market.
We customize reverse engineering and static analysis techniques to overcome challenges arising from apps' dependency on game engines~(\textbf{Section~\ref{sec:vr_privacy_compliance}}). 
Specifically, unlike prior studies that focused only on one platform, we propose a reverse engineering technique that recovers semantic information from \emph{.apk} files compiled using different game engines (e.g., Unity and Unreal) and compilation methods (e.g., IL2CPP, Mono). Recovering this information is crucial for identifying actual API usage and behaviors, enabling a more comprehensive analysis of unique compliance issues in VR applications.


\paragraph{Key Findings}
Our study reveals the \emph{status quo} of the privacy practices in the VR app ecosystem for the first time.
Some main findings are summarized below.
\begin{itemize}

    \item \textbf{Lack of a stringent privacy vetting process in different VR app stores}. VR app developers are not provided with a standardized process for disclosing privacy-related information and privacy policies. This has resulted in significant discrepancies in mandated disclosure and quality control of privacy-related declarations across VR app stores.

    \item \textbf{Low quality of privacy policy documents}. Privacy policies of existing VR apps exhibit several critical deficiencies, including insufficient detail~(30\%), inaccessible references~(20\%), limited non-English language support, lack of VR-specific content~(34\%), poor readability, and incompleteness. Such inadequate privacy policies may discourage users from adopting VR and hinder the healthy development of the VR ecosystem.


     \item \textbf{Inconsistency between behavioral profiles and declarative profiles}. In the Android-based VR app stores (Oculus and Pico), almost one-fifth of VR apps access VR-specific sensitive data without transparently and correctly declaring this behavior in their privacy policies.
\end{itemize} 
 
\paragraph{Contributions}
The main contributions of this work are summarized as follows.
\begin{itemize}
    \item \textbf{Revealing the \emph{status quo} of privacy practices in the VR app ecosystem}.  
    We conduct the first comprehensive analysis of VR apps multi-stores, revealing findings that underscore the ongoing concerns regarding the current status of the VR app ecosystem. 
    Our work serves as a call to privacy attention for this emerging and expanding domain. It encourages app developers and store operators to take measures to foster the robust advancement of the ecosystem.
    
    \item \textbf{Constructing the first multi-store dataset of VR apps for privacy analysis}.  
    We construct a large-scale VR app dataset including \appnum apps for \storenum major VR stores, and their artifacts of \apknum ~\emph{.apk} files, \webpagenum app profile pages, and \privacypolicynum privacy policies.
    

     \item \textbf{An adaptive VR app analysis framework suitable for large-scale analysis}.  
     To address the challenge posed by different store policies, diverse compilation methods, and multiple game engines in the VR domain, we extract VR-specific data APIs from relevant documentation. We then combine these with the VR corpus we collect to develop an adaptive framework that allows analyzing VR applications in different app stores. 
\end{itemize}


%
\section{Background}
In this section, we first introduce the privacy features of VR apps~(Section~\ref{sec:operating}). 
To facilitate the understanding of our app behavior analysis, we also introduce the game engines for VR app development and execution~(Section~\ref{sec:engine}).

%

\subsection{Privacy Features of VR Apps}\label{sec:operating}
Apps developed for various VR stores offer users an immersive and experiential means to engage with virtual reality content through specialized equipment like headsets, headphones, or masks. 
They often involve sensitive data to create a profound, immersive VR experience. 
For instance, some apps~\cite{menges2019improving,stein2022eye} use iris data to enable their interaction based on eye tracking, and several game or exercise apps gather and model the surrounding physical space to customize their gesture interaction. 

VR manufacturers opt for Android as their primary development system, to enable developers to create apps efficiently, such as the Oculus Quest and Pico Neo series. 
Therefore, Android-based apps are the main target in our analysis. 
Their devices naturally inherit Android's permission system as their resource access control system for sensitive resources like cameras, sensors, location services, and microphones. 
It mandates apps to explicitly declare their permission requests in the \texttt{AndroidManifest.xml} file and obtain consent from users.

\subsection{Game Engines} \label{sec:engine}
The game engines are specialized software frameworks designed to create virtual reality experiences. 
These engines offer a suite of tools and features that enable developers to construct immersive 3D environments and interactive experiences in their apps. 
They also manage device sources~(e.g., sensors and cameras), and provide program interfaces for apps to access these resources. 
Therefore, when analyzing the privacy behaviors of VR apps, we have to locate such API invocations. 

So far, two game engines, Unity and Unreal, have been dominantly popular, which together account for 70\% of the VR app engine market~\cite{singh23unreal}. 
They are both renowned for their multi-store support, enabling developers to create apps compatible with various devices. 
Both engines manage the interaction with the underlying Android OS, eliminating the necessity to develop apps in Java. 
Instead, the app crafted using them is compiled into native code and packaged into an \emph{.apk} file. 
Below, we present a brief introduction to each of the engines.


\paragraph{Unity}
Unity is the most popular game engine, according to its recent market share~\cite{unity2024market}. 
It uses {C\#}~(built on the .NET framework) for scripting, providing compatibility with a variety of stores. 
In addition to the robust graphics rendering and physics simulation, Unity offers a vast array of APIs for data handling, such as accessing the device's acceleration~(via the \texttt{Input.acceleration} API) and rotation~(via the \texttt{Input.gyro}) in 3D space. 


To analyze Unity-based apps, a challenge to address is Unity's \emph{Intermediate Language to C++} (IL2CPP) mechanism~\cite{IL2CPP}. 
For devices without the Mono runtime~\cite{mono}, IL2CPP generates {C++} code during the app building process, and compiles it into native code. 
This process introduces a challenge for app analysis, resulting in the loss of high-level information~(see Section~\ref{sec:unity}).

\paragraph{Unreal}
The Unreal engine is well known for its high-fidelity graphics and real-time rendering capabilities. 
It uses {C++} as the programming language for development, offering extensive control over game mechanics and visuals. 
Besides its advanced rendering engine that supports detailed graphic processing, Unreal supports personalized physical simulations
based on user data such as eye and body movements.
When Unreal packs the app into \emph{.apk}, its source code is compiled together with the library files into \texttt{libUE4.so}, and its main configurations are wrapped into a \emph{.pak} file.
Therefore, we mainly focus on these files for the app's behaviors in our analysis~(see Section~\ref{sec:unreal}).

\section{Methodology Overview and Domain Knowledge Aware Analysis}\label{sec:privacy_concern}

To address the challenge of performing domain knowledge-aware analysis on real-world applications, we begin by understanding the unique VR privacy threats and concerns. Then we design a methodology to measure the privacy of the VR app ecosystems. Figure~\ref{fig:vr_overview} shows the workflow of our methodology. More specifically, We first collect app data from five major VR stores. For the textual artifacts of VR apps, we analyze the disclosure of privacy-relevant information and privacy policies in different stores to answer \emph{\textbf{RQ1}}. The details are shown in Section~\ref{pp_analysis}. 
Next, we conduct an in-depth analysis of declarative documents, focusing on the provisions for protecting children's data, and cross-store permission misuse. This analysis aims to explain \emph{\textbf{RQ2}}.
For \emph{\textbf{RQ3}}, We design a static analysis framework that extracts VR-specific privacy-sensitive data APIs from apps written with Unity and Unreal game engines, compiled using different approaches, and then constructs code representations as control flow graphs.
We then combine the resulting control flow graph with the data types extracted from privacy policies to perform privacy compliance checking. 
It should be noted that our static analysis of game engines is limited to free-download VR apps (802 in total) available on the Android operating system's VR app stores, specifically Oculus and Pico.
We provide a detailed explanation in Section~\ref{sec:vr_privacy_compliance}.
\begin{figure*}[t]
    \centering
    \includegraphics[width=1\textwidth]{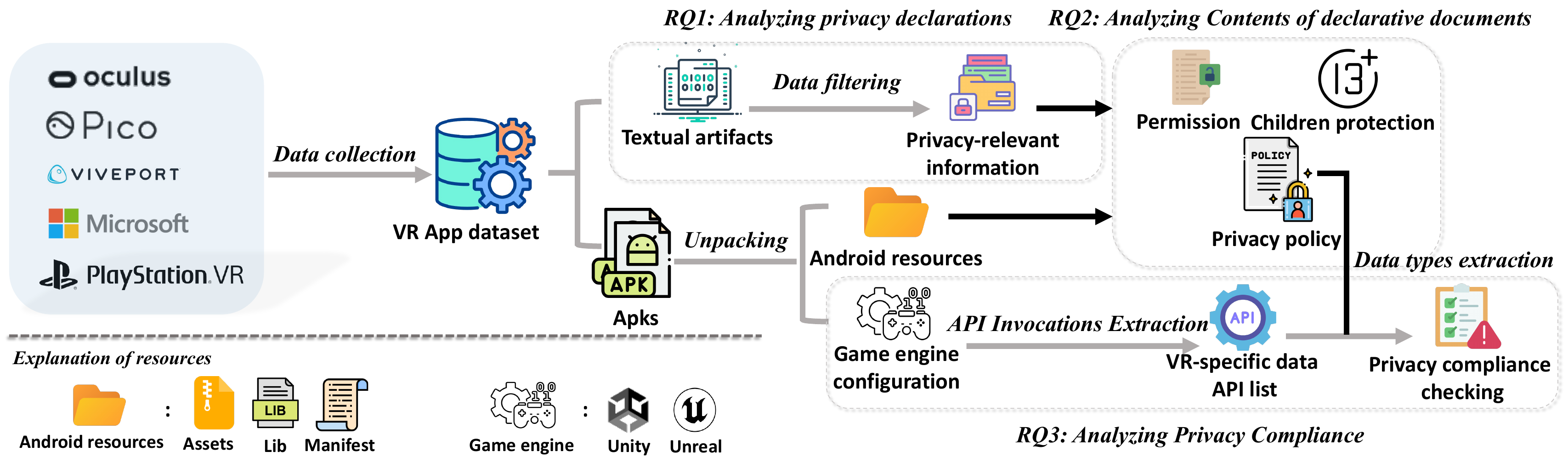}
    \caption{The workflow of privacy measurement of virtual reality app ecosystem}
    \label{fig:vr_overview}
\end{figure*}

\subsection{VR-domain Privacy-Sensitive Data and Privacy Concerns}
Since there is a lack of a comprehensive list of VR domain-specific privacy-sensitive data and privacy concerns of VR apps for our assessment use, we construct a privacy taxonomy for VR. This taxonomy is essential for guiding the design of our assessment methodology, ensuring that all relevant privacy aspects are thoroughly considered and addressed.
We resort to the literature and data regulations like GDPR for this purpose.
We start with Fernandez et al.~\cite{fernandez2022life} and Wu et al.~\cite{wu2023privacy}, which are the most recent publications reflecting VR privacy, and track the studies they cite, continuing the process. 
The referenced literature covers a wide range of topics, including VR privacy policy~\cite{trimananda2022ovrseen,ginosar2017analytical}, VR interaction~\cite{buck2021privacy}, data gathering~\cite{happa2021privacy, o2016convergence,fernandez2022life}, biometric authentication~\cite{nair2023unique,miller2020personal} and context-aware computing~\cite{buck2021privacy,fernandez2022life,trimananda2022ovrseen}.
From them, we identify \emph{privacy-sensitive Data of VR Domain}~(Section~\ref{sec:spe_data}), and consolidate privacy concerns on \emph{declarative profiles} and \emph{behavioral profiles} of VR apps~(Section~\ref{sec:concerns}). 
The overall results are summarized in Table~\ref{tab:liter_review}, and next we detail our consolidation. 

\begin{table}[t]
\centering
\setlength{\tabcolsep}{2pt}
\caption{\label{tab:liter_review}VR privacy concerns consolidated from literature}
\small
\begin{tabular}{p{1.4cm}|p{0.5cm}|p{3cm}|p{2.8cm}}
\hline

\hline
    \textbf{Concern Type}  & \textbf{No.}  &\textbf{Privacy Concern}                    & \textbf{Referred by Literature}                                   \\ \hline
\multirow{3}{*}[-2ex]{\makecell{Declarative \\ Concerns}} & C\#1      & Privacy Policy       &\makecell[lt]{~\cite{lim2022mine},~\cite{biener2022quantifying},~\cite{buck2021privacy},~\cite{happa2021privacy} \\~\cite{huang2021systematic},~\cite{kim2023facial},~\cite{lee2022technology},~\cite{lopez2016method}}
                              \\ \cline{2-4}
    & C\#2& Protection for Children   & \makecell[lt]{~\cite{spiegel2018ethics},~\cite{plechata2019age},~\cite{maloney2020anonymity},~\cite{o2016convergence}}
                               \\ \cline{2-4}  
                                                                                          
    &    C\#3        & Permission Request                       & \makecell[lt]{~\cite{adams2018ethics},~\cite{wu2023privacy}} 
 \\ \hline
  \multirow{2}{*}[-0.5ex]{\makecell{Behavioral \\ Concerns}}   &        C\#4       & Data Access                         
& \makecell[lt]{~\cite{fernandez2022life},~\cite{NairGSO23Exploring},~\cite{di2021metaverse},~\cite{heruatmadja2023biometric}}
                                    \\ \cline{2-4}
   &C\#5  & Behavior-declaration  
&\makecell[lt]{~\cite{xie2022scrutinizing},~\cite{samarin2023lessons},~\cite{liu2022evaluating}}
                               \\ \hline                                

\hline
\end{tabular}
\end{table}

\subsubsection{Privacy-sensitive Data of VR Domain}\label{sec:spe_data} 
We first summarize the data types that are treated as privacy-sensitive data in studies relevant to privacy data analysis~\cite{nair2023unique,miller2020personal,buck2021privacy,fernandez2022life,trimananda2022ovrseen}. 
Two categories of data have been the main concerns of these studies. 

\paragraph{Biometric data} 
VR apps have the capability to gather a wide spectrum of biometric data from users, including \emph{eye movements, hand gestures, body motion}, and \emph{facial expressions}.  
On the one hand, biometric data is essential for VR's user experience enhancement. 
For example, by tracking eye movements, VR apps can understand the user's focus and adjust the visual experience accordingly. 
Hand gesture tracking is crucial for enabling intuitive and natural interactions, enhancing the sense of immersion and engagement in the virtual world.

On the other hand, biometric data is typical personally identifiable information~(PII), as it is highly personalized and can be used to identify an individual uniquely. 
A recent study~\cite{nair2023unique} demonstrates that VR users can be uniquely and reliably identified across multiple sessions using just their head and hand motions.
With a larger scope of data, an accuracy of 95.3\%  can be achieved on user identification among more than 500 individuals~\cite{miller2020personal}. 

\paragraph{Ambient data} The data collected from a user's surrounding environment serves essential purposes within the realm of VR technology. 
By accurately mapping and modeling the physical space around the user, VR technology can establish virtual boundaries that correspond to real-world limitations, ensuring that users can move and interact without any physical constraints. 
Nonetheless, ambient data reveals key aspects of a person's life, such as frequently visited locations, daily routines, or social interaction, as demonstrated in previous studies~\cite{buck2021privacy,fernandez2022life,trimananda2022ovrseen}.

\subsubsection{Privacy Concerns of VR Apps}\label{sec:concerns} 
To recognize the privacy concerns of VR apps, we examine the topics of all collected papers and summarize the subjects their analyses are conducted on. 
We also read the main articles of GDPR, CCPA and Children’s Online Privacy Protection Act~(COPPA) to identify the privacy-related items that they specifically highlight. 
Three co-authors independently conduct the collection process, and then accumulate them via a discussion. 
By doing so, we identify three \emph{declarative concerns} (C\#1, C\#2, C\#3) to clarify data handling practices and restrictions on their user base, as well as two \emph{behavioral concerns} (C\#4, C\#5) to focus on the actual behaviors of VR apps when they handle VR-specific privacy-sensitive data.

\paragraph{Concern \#1: Privacy Policy}
While data protection regulations vary slightly across different regions, they all require developers~(i.e., \emph{data processors}) to provide transparent and explicit privacy policies for their VR applications, ensuring users understand how their biometric data and immersive experiences are being collected, processed, and protected, e.g., Art.13 in GDPR~\cite{gdpr}, and Art.1798.130 in CCPA~\cite{ccpa}. 
The common concerns regarding the privacy policies are on their availability~\cite{srinath2023privacy}, quality to be interpretable~(or readability)~\cite{becher2021law}, and alignment with regulatory requirements~(or completeness)~\cite{liu2021have}.

\paragraph{Concern \#2: Protection for Children}
Data regulations have been put in place specifically to protect children online, e.g., Children’s Online Privacy Protection Act~(COPPA) in the US~\cite{coppa} and GDPR Art.8. 
They all include stringent policies for handling children's data. 
For example, COPPA requires data handling for children under the age of 13 to be under parental control, and GDPR requires that information provided to children about data handling should be easy to understand. 
A recent study also indicates that children are more susceptible to addiction to VR technology compared to adults~\cite{maloney2020complicated}. 
Additionally, children often exhibit lower privacy awareness and may have insufficient understanding level~\cite{stoilova2020digital}.
Therefore, it is essential for VR apps to enforce appropriate age restrictions and explicitly disclose their handling of child users' data, to mitigate potential privacy risks to children users~\cite{behnke2014must}.

\paragraph{Concern \#3: Permission Requests}
Recall that in Android-based systems, permissions are taken as credentials to access data and resources~(Section~\ref{sec:operating}). 
As mentioned in Section~\ref{sec:spe_data}, compared to traditional Android apps, VR apps often utilize users' biometric data, requiring special permissions beyond the default Android system permissions, such as facial tracking permission in Pico (\texttt{com.picovr.permission.} \texttt{FACE\_TRACKING}) and gesture recognition permission in Oculus (\texttt{com.oculus.permission.HAND\_TRACKING}) in VR devices. These special permissions are typically platform-specific. Mixing permissions in different platforms can lead to compatibility and security risks.





\paragraph{Concern \#4: Data Access} 
The primary focus in the studies of app behaviors lies on data access and data collection~\cite{nusrat2021developers,trimananda2022ovrseen}. 
They examine the code~\cite{nusrat2021developers} or the network traffic~\cite{trimananda2022ovrseen} to check whether the app collects and transmits sensitive data to remote servers. 
Following them, we put the focus of this work on the access of VR domain-specific data. 


\paragraph{Concern \#5: Behavior-declaration Compliance}
Accessing sensitive data is not necessarily a violation, as some apps may need the data to fulfill relevant functionalities.  
This perspective has garnered a consensus in other domains such as mobile and virtual personal assistants~\cite{xie2022scrutinizing,liu2022evaluating}. 
For example, a VR Yoga app that gathers body movement data should be considered benign, as long as it transparently discloses its data handling practices and adheres strictly to its privacy policy. 
Therefore, we highlight the importance of ensuring that the \emph{behaviors} of apps should comply with the \emph{declared} practices in their policies. 

\section{Data Collection and Preprocessing} \label{data_collection}
In this section, we introduce the process of collecting data relevant to building VR apps' declarative and behavioral privacy profiles, including textual artifacts and app executables~(\emph{.apk} files)~(Section~\ref{sec:data_coll}). 
We also outline the pre-processing of the collected \emph{.apk} files~(Section~\ref{sec:apk_dec}).

\subsection{VR App Data Collection} \label{sec:data_coll}
Due to the absence of a dataset encompassing VR multi-store data, we take the initiative to construct one for our analysis. 
We collect all relevant data that can reflect the declarative and behavioral aspects of the privacy practices within the app. 

\begin{table*}[t]
\centering
\def\arraystretch{2}
\setlength{\tabcolsep}{2pt}
\caption{\label{tab:vr_market} Textual data by apps in five popular VR app stores}
\footnotesize

\begin{threeparttable}
\resizebox{1\linewidth}{!}{%
\begin{tabular}{l@{\hspace{0.4cm}}l@{\hspace{0.4cm}}l@{\hspace{0.4cm}}l@{\hspace{0.4cm}}r@{\hspace{0.4cm}}r@{\hspace{0.4cm}}r@{\hspace{0.4cm}}r@{\hspace{0.4cm}}r@{\hspace{0.4cm}}r@{\hspace{0.4cm}}r@{\hspace{0.4cm}}r@{\hspace{0.4cm}}r@{\hspace{0.4cm}}r@{\hspace{0.4cm}}r@{\hspace{0.4cm}}r@{\hspace{0.4cm}}r@{\hspace{0.4cm}}r@{\hspace{0.4cm}}r@{\hspace{0.4cm}}r}
\textbf{VR App store}& \textbf{\rotatebox{90}{\makecell[lt]{Operating\\ System}}}& \textbf{\rotatebox{90}{Manufacture}}& \textbf{\rotatebox{90}{Size (Apps)}}& \textbf{\rotatebox{90}{Developer}}& \textbf{\rotatebox{90}{Copyright}}& \textbf{\rotatebox{90}{\fbox{Age Rating}}}& \textbf{\rotatebox{90}{\fbox{Category}}}& \textbf{\rotatebox{90}{\fbox{Permission}}}& \textbf{\rotatebox{90}{\fbox{Language}}}& \textbf{\rotatebox{90}{Review}}& \textbf{\rotatebox{90}{\fbox{Privacy Policy}}}& \textbf{\rotatebox{90}{Game Model\tnote{1}}}& \textbf{\rotatebox{90}{\makecell[lt]{Update\\ Information}}}& \textbf{\rotatebox{90}{App Size}}& \textbf{\rotatebox{90}{\makecell[lt]{System\\ Requirement}}}& \textbf{\rotatebox{90}{\fbox{\makecell[c]{Environmental\\Requirement\tnote{2}}}}}& \textbf{\rotatebox{90}{\makecell[lt]{Device\\ Requirement}}}& \textbf{\rotatebox{90}{\fbox{Play Styles\tnote{3}}}}& \textbf{\rotatebox{90}{\makecell[c]{Supported \\Controllers\tnote{4}}}}                  \\ \hline

\hline
Oculus      &Android &Meta &  2,233 & \ding{51}&                      & \ding{51}             & \ding{51} &                   & \ding{51} & \ding{51}    &\ding{51}             &\ding{51} &\ding{51}  &\ding{51} &          &      &\ding{51} &\ding{51}  & \ding{51}                                                                                     \\ \hline
Viveport    &Android/Windows&HTC&  3,103 &\ding{51} &                      & \ding{51}            & \ding{51} &                    & \ding{51} & \ding{51}            & P$^\dagger$(274)           &\ding{51}          &\ding{51}  &\ding{51}          &\ding{51} &\ding{51}  &\ding{51} & &\ding{51}                                                                                   \\ \hline
Pico        &Android &Pico Interactive& 455 & \ding{51}&                      & \ding{51}            & \ding{51} & \ding{51}          & \ding{51} & \ding{51}             & \ding{51}            &\ding{51} &\ding{51}  &\ding{51} &          &       &\ding{51} & \ding{51}  & \ding{51}                                                                           \\ \hline\hline
Microsoft   &Windows &Microsoft& 473&\ding{51} &P$^\dagger$(157)              & \ding{51}            & \ding{51} &           & \ding{51} & P$^\dagger$(83)  &P$^\dagger$(206)           &\ding{51}          &           &\ding{51} &P$^\dagger$(451) &     &\ding{51}  &  &                                                                                        \\ \hline
PlayStation &Psos &Sony& 301 &\ding{51} &\ding{51}           & \ding{51} & \ding{51}          &           & \ding{51} &              & P$^\dagger$(33)            &\ding{51} &           &          &\ding{51} &         &\ding{51}  & \ding{51} &                                                                           \\ \hline

\hline

\end{tabular}}
\begin{tablenotes}
\item[1] Game Model: The number of users supported by the VR apps and whether it can be networked to interact with other users.
\item[2] Environment Requirement: Venue requirements for VR apps, such as room size and ground conditions.
\item[3] Play Styles: VR apps support the way users use them. such as standing, sitting, or lying.
\item[4] Supported Controllers: VR apps support controllers such as joysticks, steering wheels and more.
\end{tablenotes}
\begin{flushleft}
    \textsuperscript{$\dagger$} P: Partial, indicating that part of apps make the disclosure. The boxed \fbox{features} stand for those relevant to apps' privacy profiles.
\end{flushleft}

\end{threeparttable}
\vspace{-0.2cm}
\end{table*}

\paragraph{Textual artifacts} App stores typically require developers to provide meta-information for their apps, as exemplified in Figure~\ref{fig:vr_example}. 
This encapsulates the declarative aspect of the app's privacy practices, encompassing details such as the link to the privacy policy, requested permissions, age rating, and interaction styles. 
Hence, we create a crawler based on Selenium~\cite{Selenium} for the data collection and parsing. 
The process starts at the portal page of the app catalog, systematically navigating through all available pages and extracting the textual content from each app's profile page. This method allowed us to compile a comprehensive dataset, resulting in the collection of textual artifacts from a total of 6,565 VR apps. The details and analysis of this dataset are summarized in Table~\ref{tab:vr_market}.

\begin{figure}[t]
    \centering
    \includegraphics[width=0.45\textwidth]{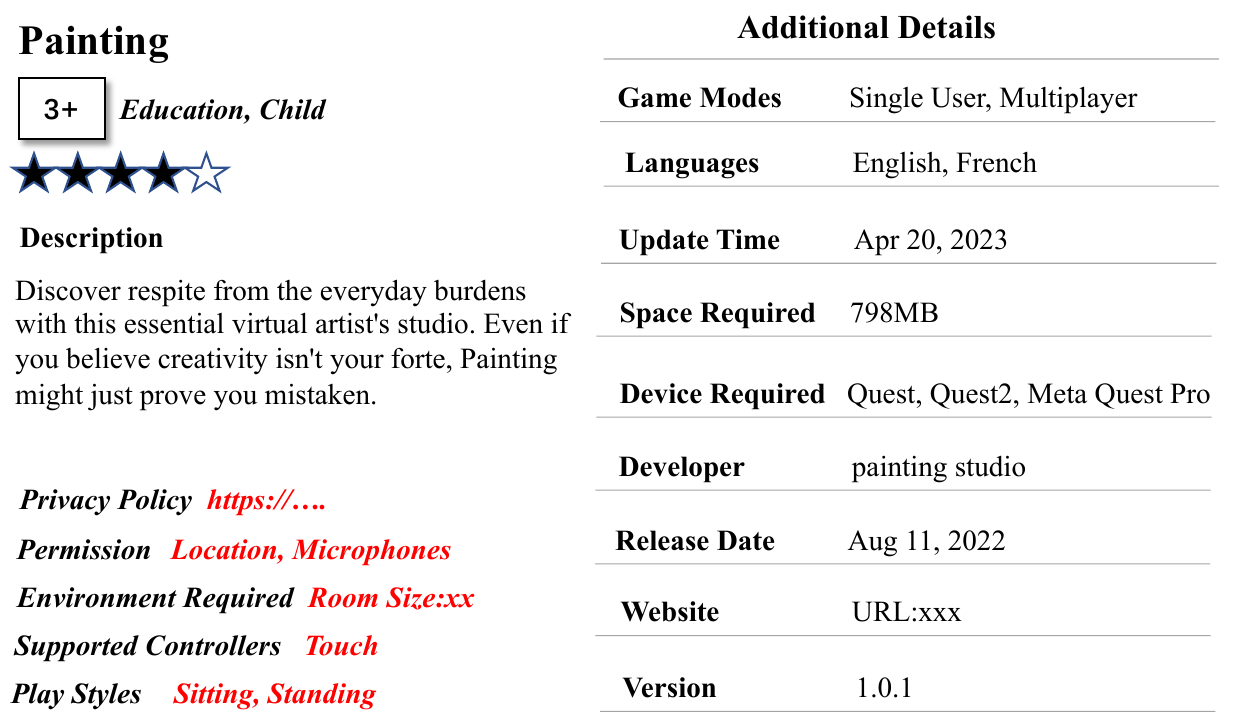}
    \caption{An example of textual artifacts of a VR app \textnormal{(with privacy-related texts highlighted\protect\footnotemark)}}
    \label{fig:vr_example}
    \vspace{-0.2cm}
\end{figure}
 \footnotetext[1]{We consider \emph{Environment Required} and \emph{Play Styles} privacy-relevant, as when an app enforces requirements on ambiance and the interaction model, it usually checks using a camera upon being started}

\paragraph{App executables}
Given that Android-based apps are more accessible for analysis than apps for other stores, we mainly collect apps' \emph{.apk} files for subsequent study. 
We first create multiple accounts on app stores and deploy them on six VR devices, including three Meta Quest 2, two Pico Neo 3, and one Pico 4.
From the app stores, we install all free-to-use apps onto the devices. 
Then, we sideload the APK Extractor~\cite{apkextractor} on the devices, which is a tool designed for extracting \emph{.apk} files installed on Android devices. 
Finally, we use the Android Debug Bridge (ADB) \texttt{pull} command to transfer these extracted \emph{.apk} files to a PC. 
Among the five targeted VR stores, three of them, i.e., Oculus, Viveport, and Pico, primarily use Android as their operating system. 
In total, we have collected \apknum \emph{.apk} files, including \oculusapknum from Oculus and \picoapknum from Pico. 
As Viveport has a limited number of free-to-use apps provided, and most of them are for PC  (\emph{.exe} files), apps from its app store are not included in our dataset.

\subsection{APK Unpacking} \label{sec:apk_dec}
After obtaining the \emph{.apk} files, we unpack them to extract resources that our assessment relies on. 
We use the ApkTool~\cite{apktool} which has been widely used for reverse engineering and analyzing Android-based applications. 
The \emph{.apk} files are built by the development engines, such that their internal structures are organized differently. 
We extract the following three key resources for subsequent analysis. 

\paragraph{Assets}
In traditional Android apps, the assets folder is used for storing resource files like text, audio, video, or configuration files that do not require compilation. 
In Unity-based apps, Unity uses it to store game models, bytecode and native code. 
Particularly, it stores bytecode~(compiled from C\# scripts) for apps targeting Mono runtime, and native code for apps running without interpreters. 
The extracted code is used for the app behavior analysis~(Section~\ref{sec:unity}). 

\paragraph{lib}
The lib directory in \emph{.apk} is designated for storing an app's native library files. 
In traditional Android apps, these libraries are typically written in low-level programming languages like C or C++, and are compiled into native code. 
It is utilized by Unreal to store the entire app developed based on it, the configurations of the app, and the Unreal runtime. 
All code and runtime are compiled into a library named \texttt{libUE4.so}, and all configurations are wrapped into a \emph{.pak} file. 
Both are located in the lib directory.
We mainly focus on these files for the app's behaviors in our analysis~(Section~\ref{sec:unreal}).

\paragraph{AndroidManifest}
The \texttt{AndroidManifest.xml} file is a crucial configuration file that provides information about the app to Android OS, such as package information, application components, and intent filters. 
It is used by Android OS to understand the structure and behavior of the app at the installation time. 
In our analysis, we use the permission requests that are contained in this file~(Section~\ref{sec:permission}). 

%
\section{Analyzing Privacy Declarations (RQ1)}\label{pp_analysis}
%

To address the question of how well VR apps provide users with adequate documentation on their data handling practices, we analyze the status of apps' declarative documents including privacy-relevant information and the privacy policy (\textbf{Concern \#1}) among five app stores.  
\subsection{Overview of Privacy-relevant Information Disclosure}\label{sec:cross_store}
As shown in Table~\ref{tab:vr_market}, there is a significant variation in the data types of apps disclosed among app stores. 
Surprisingly, except for Pico, the other two Android-based app stores do not mandate the necessary app permissions in their textual data.
On the other hand, the situation with privacy policies is also concerning. 
Apart from Oculus and Pico, apps on the other three stores often lack privacy policies, with specific instances being 56.45\% on Microsoft, 89.04\% on PlayStation, and 91.17\% on Viveport.

We also treat the external controllers as privacy-relevant information, as they may involve gesture movement of the user. 
For this type of information, Microsoft and PlayStation do not mandate the disclosure. 
Interaction models and environmental requirements should be disclosed, too, as apps require them to conduct checking through cameras when apps are started. 
Pico and PlayStation are the only two stores that mandate the disclosure of the interaction model, and Viveport mandates the environmental requirements.  

\begin{formal}

\noindent \textbf{Finding \thefinding: \stepcounter{finding}There is a lack of standard policy on the disclosure of privacy-relevant features by VR apps} 
The mandated information for disclosure exhibits significant discrepancies among VR app stores. In comparison, Oculus and Pico provide more comprehensive privacy-relevant information. This is not only because they focus exclusively on VR app stores, but also due to the benefits of the complete Android ecosystem. The current lack of comprehensive information disclosure in multiple VR stores has raised concerns about the transparency and accuracy of user privacy data.
\end{formal}
\subsection{Analysis of VR Privacy Policies}\label{sec:ava}
Previous research~\cite{garrido2023sok} highlights that compared to mobile apps, VR apps are more likely to collect sensitive user data. If these practices are not clearly disclosed in the privacy policy, it can lead to significant privacy risks, including unauthorized data use, potential breaches, and a lack of informed consent from users, ultimately compromising user trust and regulatory compliance.

\begin{table}[t]
\small
\caption{The distribution of (failure) status codes from links in privacy policies among app stores\label{status_code}}
\scalebox{0.76}{
    \begin{tabular}{l@{\hspace{0.01cm}}r@{\hspace{0.8cm}}r@{\hspace{0.8cm}}r@{\hspace{0.8cm}}r@{\hspace{0.8cm}}r}
    \toprule[1pt]
     & \textbf{Oculus} & \textbf{Viveport} & \textbf{Pico} & \textbf{Microsoft} & \textbf{PlayStation}\\
    \hline \textbf{404 (\%)} & 1.7 & 8.8 & 21.8 & 10.2 & 5.9\\
    \textbf{400 (\%)}& 0.0 & 0.4 & 0.0 & 0.5 & 0.0 \\
    \textbf{500 (\%)} & 0.0 & 1.1 & 0.7 & 1.0 & 0.0  \\
    \textbf{503 (\%)} & 0.1 & 0.0 & 0.3 & 0.0 & 0.0  \\
    \textbf{410 (\%)} & 0.2 & 0.0 & 0.0 & 0.0 & 0.0  \\
    \textbf{Timeout (\%)} & 3.4 & 11.0 & 9.1 & 11.7 & 2.9  \\
    \hline \textbf{Total (\%)} & 5.4 & 21.3 & 31.9 &23.4 &8.8 \\
    \bottomrule[1pt]
    \end{tabular}
    
}

\end{table}


\paragraph{Validity of privacy policies}
Some apps provide non-informative privacy policies. For instance, a few privacy policies consist of only a single sentence, such as ``\emph{We will collect your data}''. Such simplistic sentences fail to inform users about how their data is collected and processed by the data handlers, especially regarding VR-specific data like eye-tracking and body movement, which are closely tied to user identity. As discussed in~\cite{meier2020shorter}, an overly short privacy policy fails to provide sufficient information. 
We treat policies shorter than 100 words as insufficient~\cite{xie2022scrutinizing}.

 

On the other hand, we find that some apps~(3.87\%) provide the link to the app store's privacy policy as the policy for their apps. 
These policies broadly summarize the data handling practices of official applications provided by device manufacturers but often express data collection practices in obscure language qualifiers, such as ``\emph{We \underline{may} collect your biometric data}''. However, this ambiguity fails to clearly inform users whether and what specific VR-related biometric data is being collected. Since VR-collected data can introduce unique threats, failure to inform these threats can raise concerns about fairness to users and is explicitly prohibited by Article 29 Working Party~\cite{ARTICLE29}.



\begin{formal}

\noindent \textbf{Finding \thefinding: \stepcounter{finding} 29.8\%  apps fail to provide informative privacy policies.} 
In each store, there are apps that fail to provide informative privacy policies. 
Apps in the Oculus store perform slightly better due to Meta's strict policies for developers. However, it is still notable that over 20\% of the apps have privacy policies that are less than 100 words long.
Some apps simply provide the privacy policies of their app store as their own policies. 
All stores except PlayStation suffer from this issue, including Microsoft (12.68\%), Viveport (3.28\%), Oculus (2.46\%), and Pico (7.46\%). 
    
\end{formal}
\vspace{5pt}

\paragraph{Accessibility of Links}
VR apps' privacy policies often reference other web pages for explanatory purposes, such as defining VR-specific terms or explaining data processing methods unique to virtual environments. To ensure privacy policies are clear and transparent, these links must stay functional and accessible, enabling users to understand data collection related to VR-specific features. We check the links and record the returned status codes. The distribution of inaccessible links across five VR app stores is detailed in Table~\ref{status_code}.

\begin{formal}
\noindent \textbf{Finding \thefinding: \stepcounter{finding} Privacy policies of existing VR apps largely suffer from the inaccessibility of referred resources. }
More than 20\% VR apps fail to maintain the accessibility of the resources their privacy policies refer to. 
The most severe issues stem from a status code of 404~(5.69\%) and access timeouts~(5.34\%). 
Among app stores, Pico exhibits the worst inaccessibility, where over 30\% of privacy policy links cannot be accessed properly, largely due to Pico's lack of supervision over apps.
\end{formal}
\vspace{5pt}

\paragraph{Language Support}
In traditional app stores, most apps primarily support English, largely because the companies and developers are predominantly based in English-speaking countries. In contrast, with the advent of open-source game engines like Unreal, the development of VR app stores has become more diverse. For example, China's Pico and Japan's PlayStation VR store cater to a global audience. Developers from around the world can now upload their VR apps to different app stores, making them accessible to users worldwide.
Hence, for apps that support multiple languages, we also consider the coverage of the multilingual versions as an important metric in the availability of privacy policies, as Article 29 Working Party~\cite{ARTICLE29} stipulates that when the target data subjects of a data collector use multiple languages, corresponding translated versions should be provided. 
We thus check if the language versions of the app's privacy policy match the list of language versions it claims to support.

We build a language detector based on Selenium~\cite{Selenium} for this purpose. 
It detects whether a language selection option is available on the webpage of the privacy policy by searching for buttons and links using keywords of ``\emph{language}'' and language names like ``\emph{English}'' and ``\emph{French}''. 
If such a link exists, it simulates a click action and checks its availability. 
It also mutates the language code in the privacy policy link, such as ``\emph{fr}'' for French and ``\emph{de}'' for German, and checks whether the mutants lead to accessible privacy policies. 
After obtaining these pages, it filters out invalid policies and uses \emph{langdetect}~\cite{langdetect} to check the languages of the pages.  

\begin{figure}[t]
    \centering
    \includegraphics[width=0.5\textwidth]{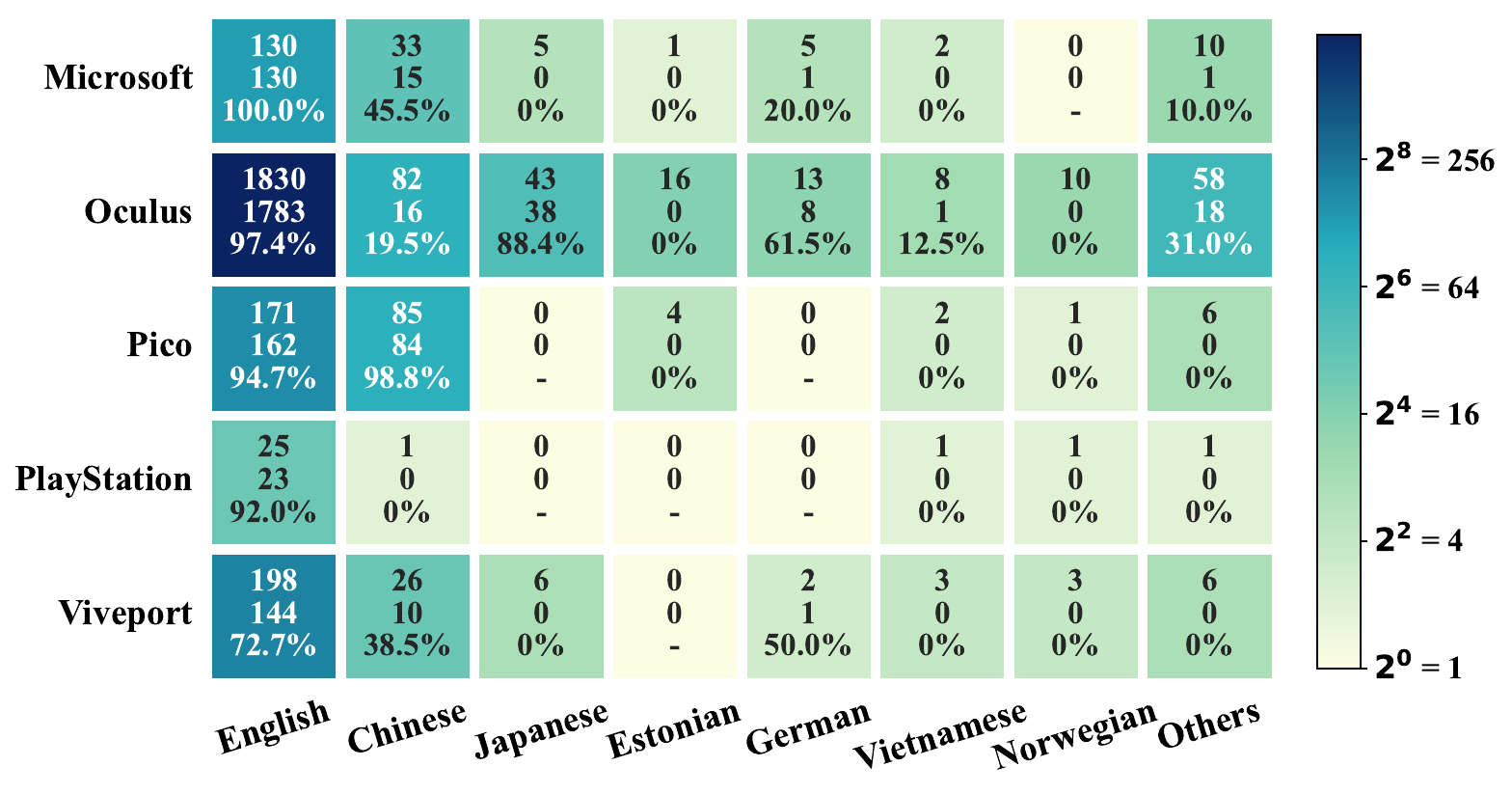}
    \caption{The coverage of multilingual versions of privacy policies across VR stores\protect\footnotemark}
    \label{fig:language}
    \vspace{-0.4cm}
\end{figure}

\footnotetext[2]{The first row of the cell represents the number of privacy policies in the corresponding language, the second row indicates the number of apps whose supported languages match their privacy policies, and the third row shows the proportion.}
Our language detector helps us identify the presence of many privacy policies in languages other than English, including Chinese~(309, 11.08\%), Japanese~(86, 3.08\%), German~(25, 0.90\%), Vietnamese~(20, 0.72\%), French~(15, 0.54\%) and other languages~(97, 3.48\%).
Figure~\ref{fig:language} summarizes apps' coverage of multilingual versions of their privacy policies.

\begin{formal}
\noindent \textbf{Finding \thefinding: \stepcounter{finding} Existing VR apps perform well in covering English privacy policy, but fall short in covering non-English languages as they claim in their ``\emph{Supported Language}'' sections.}
\end{formal}

\paragraph{VR Specificity}
Some developers simply reuse the privacy policy for their other apps as the policy for their VR apps. 
Such policies only provide general information to users, without important details of how data is processed specifically in VR apps. 
This behavior also increases the risk of non-compliance with privacy laws and regulations, potentially exposing both developers and VR app stores to legal and security issues.
To check the specificity, we search the policy for the name of the VR app and VR-related terms, such as ``\emph{virtual reality}'', ``\emph{VR}'', ``\emph{immersive}'', and ``\emph{head-mounted display}''. 
We conduct a manual check on the identified cases for confirmation and find that most cases are with PlayStation and Microsoft, and those privacy policies are for console and PC games. 
This is likely because both stores develop console and PC games before venturing into VR. 
\begin{formal}
\noindent \textbf{Finding \thefinding: \stepcounter{finding} Almost 1/3 privacy policies are not specific to their VR apps.} 
Existing VR apps significantly suffer from this issue, with 92.67\% in PlayStation, 71.47\% in Microsoft, 68.64\% in Viveport, 29.27\% in Pico and 18.61\% in Oculus. The reason for this situation is that before developing VR app stores, PlayStation, Microsoft, and Viveport had other app stores such as console games and PC apps. Therefore, many developers simply transfer the privacy policies from their previous apps to VR apps. Pico and Oculus, only provide VR app stores, which results in better performance in this regard.
\end{formal}

We also analyze the readability of the privacy policies across apps in five VR app stores. The details are provided in Appendix~\ref{sec:quality}.


%
\section{Analyzing Contents of Declarative Documents~(RQ2)}\label{sec:doc_content}

In this section, we examine the content of VR apps' declarative documents from two angles, focusing on the consistency of age restrictions for child users (\textbf{Concern \#2}) and cross-store permission misuse (\textbf{Concern \#3}), to address RQ2. 

\subsection{Protection for Children}\label{age}

\begin{table*}[t]
\caption{The status of child-related privacy policies across app categories\protect\footnotemark}\label{age_check}
\small
\resizebox{1\linewidth}{!}{%
\begin{tabular}{c|c c c c c|c c c c c|c c c c c|c c c c c|c c c c c}
\hline

\hline \multirow{2}{*}{ \textbf{Category} } & \multicolumn{5}{c|}{ \textbf{Oculus} } & \multicolumn{5}{c|}{ \textbf{Viveport} } & \multicolumn{5}{c|}{ \textbf{Pico} } & \multicolumn{5}{c|}{\textbf{Microsoft}}  & \multicolumn{5}{c}{\textbf{PlayStation}}\\ \cline{2-26}
 & \multicolumn{1}{c|}{App} & \multicolumn{1}{c|}{I} & \multicolumn{1}{c|}{\%} & \multicolumn{1}{c|}{D}& \multicolumn{1}{c|}{\%} & \multicolumn{1}{c|}{App} & \multicolumn{1}{c|}{I} & \multicolumn{1}{c|}{\%} & \multicolumn{1}{c|}{D} & \multicolumn{1}{c|}{\%} & \multicolumn{1}{c|}{App} & \multicolumn{1}{c|}{I} & \multicolumn{1}{c|}{\%} & \multicolumn{1}{c|}{D} & \multicolumn{1}{c|}{\%}
 & \multicolumn{1}{c|}{App} & \multicolumn{1}{c|}{I} & \multicolumn{1}{c|}{\%} & \multicolumn{1}{c|}{D} & \multicolumn{1}{c|}{\%} & \multicolumn{1}{c|}{App} & \multicolumn{1}{c|}{I} & \multicolumn{1}{c|}{\%} & \multicolumn{1}{c|}{D} & \multicolumn{1}{c}{\%} \\ \hline
 Shooter &76 &-&-&4 &5.3 &90 &40 &44.4 &0  &0.0 &51 &5 &9.8 &5  &9.8 &12 &6 &50.0 &0  &0.0 &22 &18 &81.8 &0  &0.0\\ \hline
 Action  &646 &- &- &49 &7.6 &326 &144 &44.2 &0 &0.0 &87 &9 &10.3 &4 &4.6 &55 &27 &49.1 &1 &1.8  &81 &64 &79.0 &2 &2.5\\ \hline
 Simulation  &156 &- &- &5 &3.2 &238 &208 &87.4 &0 &0.0 &15 &2 &13.3 &1 &6.7 &18 &13 &72.2 &0 &0.0 &14 &9 &64.3 &0 &0.0 \\ \hline
 Adventure  &259 &- &- &15 &5.8 &639 &286 &44.8 &5 &0.8 &27 &1 &3.7 &2 &7.4 &- &- &- &- &- &50 &35 &70.0 &4 &8.0\\ \hline
 Horror  &60 &- &- &5 &8.3 &19 &1 &5.3 &0 &0.0 &5 &1 &20.0 &0 &0.0 &- &- &- &- &- &10 &7 &70.0 &0 &0.0\\ \hline
 Education  &190 &- &- &3 &1.6 &89 &73 &82.0 &0 &0.0 &13 &10 &76.9 &0 &0.0 &98 &95 &96.9 &0 &0.0 &- &- &- &- &- \\ \hline
 Music  &76 &- &- &2 &2.6 &36 &34 &94.4 &0 &0.0 &21 &4 &19.0 &0 &0.0 &3 &1 &33.3 &0 &0.0 &42 &39 &92.9 &0 &0.0 \\ \hline
 Social  &88 &- &- &4 &4.5 &3 &3 &100.0 &0 &0.0 &18 &11 &61.1 &0 &0.0 &- &- &- &- &- &- &- &- &- &- \\ \hline
 Puzzle  &102 &- &- &1 &1.0 &- &- &- &- &- &28 &2 &7.1 &0 &0.0 &17 &12 &70.6 &0 &0.0 &12 &8 &66.7 &1 &8.3 \\ \hline
 Sport  &151 &- &- &2 &1.3 &43 &31 &72.1 &0 &0.0 &41 &6 &14.6 &1 &2.4 &14 &13 &92.9 &0 &0.0 &8 &7 &87.5 &0 &0.0\\ \hline
 Casual  &306 &- &- &4 &1.3 &721 &584 &81.0 &2 &0.3 &51 &18 &35.3 &1 &2.0 &- &- &- &- &- &8 &5 &62.5 &0 &0.0 \\ \hline
 Strategy  &29 &- &- &1 &3.4 &11 &5 &45.5 &0 &0.0 &11 &1 &9.1 &0 &0.0 &8 &5 &62.5 &0 &0.0 &4 &4 &100.0 &0 &0.0 \\ \hline
 Media  &23 &- &- &0 &0.0 &581 &507 &87.3 &1 &0.2 &45 &35 &77.8 &0 &0.0 &10 &9 &90.0 &0 &0.0 &- &- &- &- &- \\ \hline
 Business  &20 &- &- &0 & 0.0 &26 &21 &80.8 &0 &0.0 &- &- &- &- &- &33 &24 &72.7 &0 &0.0 &- &- &- &- &-\\ \hline \hline
 \textbf{Total}  &- &- &- &- & 3.3 &- &- &66.7 &- &0.1 &- &- &27.5 &- &2.5 &- &- &69.0 &- &0.2 &- &- &77.5 &- &1.9\\ \hline  
 
 \hline


\end{tabular}}
\begin{tablenotes}
\item[1]I~(Inconsistency): the app is for general user groups, but its privacy policy does not include a children's data protection regulation.
\item[2]D~(Discrepancy): the app's age restriction exceeds the age limitation stated in its privacy policy. 
\end{tablenotes}
\end{table*}

Due to the immersive nature of VR apps, which can lead to issues such as overstimulation and reduced awareness of the real world for users~\cite{li2022towards}, especially for children who may be more vulnerable to these effects, VR stores require developers to categorize apps based on their content and apply age restrictions~\cite{vlahovic2023initiative}.
For example, a shooting game is categorized as \emph{shooter} and \emph{action}, suitable for users above 13 years old, and a drawing app is categorized as \emph{education} and \emph{casual},  suitable for all user groups. 
We explore two questions in the assessment of child user protection, including \textbf{Q1)} for apps suitable for all user groups, whether child data protection clauses are included in their privacy policies, and \textbf{Q2)} for apps with age restrictions, whether the age limitations are consistent with those stated in their privacy policies. 
An example of violating Q2 is that a VR app in the \emph{shooter} category prohibits users under the age of 13, but its privacy policy defines the age limitation as 11, causing disputes when it collects data of a 12-year-old user. 

For Q1, we use QuPer~\cite{yan2024on} to detect the component of \texttt{CHILDREN} when the app has no restriction on the user groups. 
For Q2, we first locate sentences in the \texttt{CHILDREN} section and determine whether they include numbers that signify ages. 
To do this, we employ SpaCy~\cite{spacy} to examine the tokens preceding or following the numbers for specific terms like ``\emph{years}'', ``\emph{age}'', and ``\emph{old}''.
We manually select 20 privacy policies containing the children component for confirmation, and all ages specified by them can be extracted using our method. 
The results of Q2 are summarized in Table~\ref{age_check}. 




\begin{formal}
\noindent \textbf{Finding \thefinding: \stepcounter{finding} Existing VR apps fall short in enforcing age restrictions and disclosing specific privacy policies on child users' data.}  
Only apps in Oculus have clearly defined age restrictions. 
The other four app stores have an average of 49.1\% of apps suitable for users of all age groups, but many of them fail to include specific children's privacy clauses in their privacy policies, 
particularly, 66.7\% in Viveport, 27.5\% in Pico, 69.0\% in Microsoft, and 77.6\% in PlayStation. 

\noindent \textbf{Finding \thefinding: \stepcounter{finding} The discrepancy between the age restriction disclosed through the app store and the age limitation defined in the privacy policy is not uncommon.}
The discrepancy occurs in all app stores, with Oculus (3.3\%), Pico (2.5\%) and PlayStation (1.9\%) suffering more than the other two. 
Notably, in the \emph{shooter} category of Pico and \emph{horror} category of Oculus, this discrepancy reaches 9.8\% and 8.3\%, respectively. 
\end{formal}

\subsection{Cross-store Permission Misuse}  \label{sec:permission}
The permission system plays a central role in resource access control in Android-based systems. 
The two Android-based platforms, i.e., Pico and Oculus, extend the native Android permission system with their custom permission labels for VR-specific data, 
e.g., the facial tracking permission \texttt{com.oculus.permission.FACE\_TRACKING} in Oculus, and \texttt{com.picovr.permission.FACE\_TRACKING} in Pico.
These permission labels are specific to their respective platforms, so when an app attempts to use a permission from one platform on another, the compiler may fail to match or recognize these permissions, potentially leading to compilation errors or causing the permission system to malfunction at runtime.
However, we discover that some Oculus permissions in Pico \emph{.apks}, including \texttt{HAND\_TRACKING}, \texttt{FACE\_TRACKING} and \texttt{EYE\_TRACKING}, etc.

\begin{formal}
\noindent \textbf{Finding \thefinding: \stepcounter{finding} There is a discrepancy between requested permissions and declared permissions in the Pico app store.} 
Pico outperforms Oculus by mandating the disclosure of permissions. Nonetheless, a notable proportion~(7.2\%) of VR apps on Pico have requested at least one permission that is not indicated on their homepage. 
Such permissions include \texttt{READ\_EXTERNAL\_STORAGE}, \texttt{CAMERA} and \texttt{READ\_PHONE\_STATE}. 

\noindent \textbf{Finding \thefinding: \stepcounter{finding} There are cross-platform declarations for VR-specific permissions present in Pico.} 
We discover a total of 27 Oculus VR-specific permission declarations across the \emph{.apks} of 12 Pico apps, highlighting unreasonable compatibility issues between VR platforms and the “wrapper app.” This might cause potential permission issues, and we will conduct more studies in the future to further explore this observation.
\end{formal}

We also examine the completeness of the privacy policies across apps in five VR app stores. The details are provided in Appendix~\ref{sec:comp}.

\section{Analyzing Privacy Compliance (RQ3)}\label{sec:vr_privacy_compliance}
In this section, we evaluate apps' data access~(\textbf{Concern \#4}) and behavior-policy compliance~(\textbf{Concern \#5}), by analyzing the code and privacy policies of Android VR apps.

\subsection{API Invocations Extraction}
We design different methods for extracting APIs related to VR-specific privacy-sensitive data based on different game engines and compilation modes. 
The differences in compilation language and file structure make the analysis of different VR applications very challenging. This also prompts us to find methods to analyze VR apps across different development environments. To solve this challenge, we summarize the APIs used to access VR-specific privacy-sensitive data from the official SDK documentation of Pico and Oculus. The mapping is listed in Appendix~\ref{app:map}. We take the invocations of these APIs as indicators of an app's behavior in accessing sensitive data. Moreover, by creating a large corpus of VR-specific data types using domain-specific knowledge, we design a comprehensive and adaptive analysis framework that supports compliance checking for VR applications across different stores.
In the remainder of this section, we brief the mechanisms of these two engines and outline our analysis.

\subsubsection{Unity-based Apps} \label{sec:unity}
Unity uses two modes of compiling and executing C\# scripts, i.e., \emph{Mono} and \emph{Intermediate Language to {C++} Compiler (IL2CPP)}. 
As discussed in Section~\ref{sec:apk_dec}, in the former, the bytecode can be obtained, and in the latter, only native code is available for our analysis. 
Below, we describe our approach to analyzing apps in these two scenarios. 

\footnotetext[3]{We only consider categories that are present in each VR app store and have a substantial number of apps.}

\paragraph{Mono}
Mono is an open-source implementation of the .NET Framework. 
The Mono-oriented apps are initially compiled from C\# code into a universal Intermediate Language (IL). 
At runtime, the IL code is Just-In-Time (JIT) compiled into machine code for execution. 
We employ \emph{dotPeek}~\cite{dotpeek} to translate the IL code back into C\# code, on which our static analysis is conducted. 
Next, we construct the Abstract Syntax Tree~(AST) of the code using \emph{Roslyn}~\cite{roslyn}, and traverse through its nodes to search for the sensitive API calls. 

\paragraph{IL2CPP}
IL2CPP compiles {C\#} code into {IL} code, which is then converted into {C++} code. 
The {C++} code is compiled into native binary code for the target platform, resulting in a file named \texttt{libil2cpp.so}. 
Meanwhile, the meta-information is kept in another file named \texttt{global-metadata.dat}. 
This file can be decoded with an open-source tool named {IL2CPPDumper}~\cite{IL2CPPDumper}, and the following two files can be derived. 
\begin{itemize}
    \item \texttt{dump.cs}, which is the symbol table that contains classes, functions, fields, and their offsets. 
    \item \texttt{DummyDll/Assembly-CSharp.dll}, which includes all function declarations of the main app. 
\end{itemize}

\begin{figure}[t]
    \centering
    \includegraphics[width=0.5\textwidth]{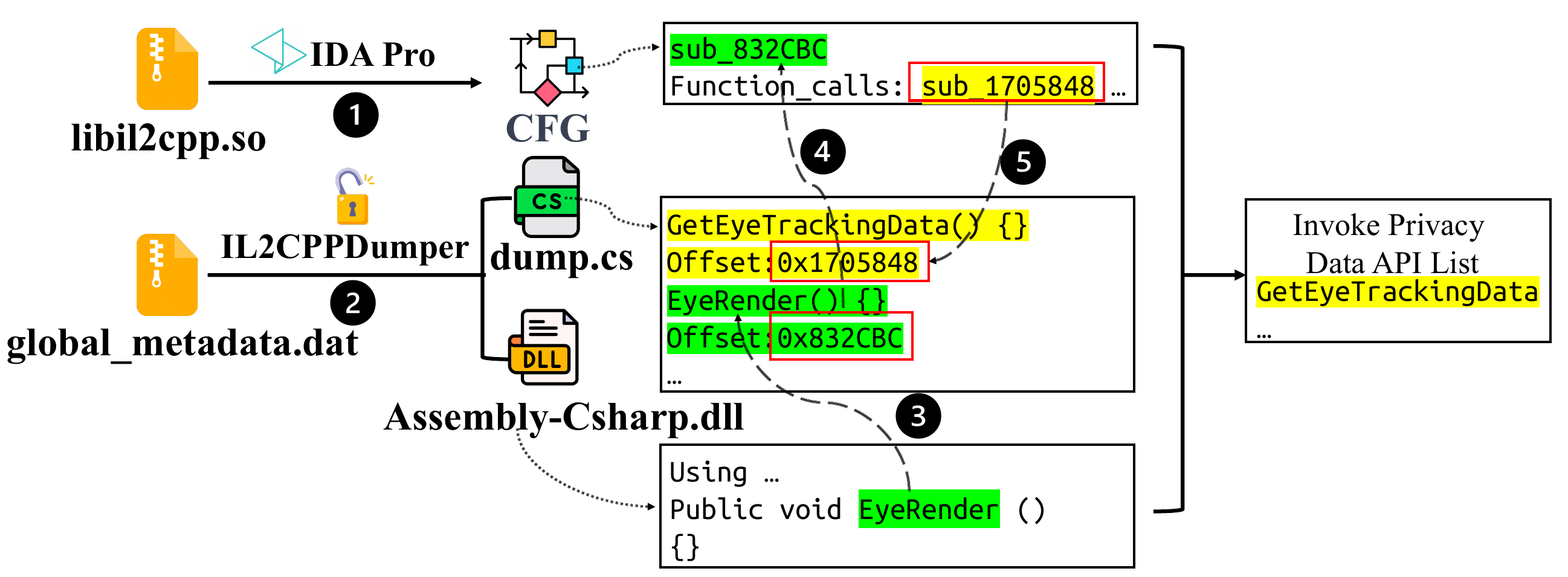}
    \caption{A demonstration of IL2CPP native code analysis}
    \label{fig:unity_static} 
\end{figure}

Figure~\ref{fig:unity_static} illustrates our approach of extracting API calls. 
We first decompile \texttt{libil2cpp.so} and construct the control flow graph~(CFG) using IDA Pro~(\ding{182}). 
An intuitive way of extracting API calls is through searching the CFG, but the CFG is too huge to accommodate this. To solve this challenge, we start with the functions declared in \texttt{Assembly-CSharp.dll}. 
Taking \texttt{EyeRender()} shown in Figure~\ref{fig:unity_static} as an example, we get its offset from the symbol table~(\ding{184}), i.e., \emph{832CBC}. 
With this offset, we locate the sub-graph of \texttt{EyeRender()} in the CFG~(\ding{185}), where we find that another function with offset \emph{1705848} is invoked. 
Through looking up the symbol table~(\ding{186}), this offset corresponds to the API \texttt{GetEyeTrackingData()} (used for collecting eye tracking data). 
By recursively executing this process, we can identify all sensitive data-related APIs invoked by the app. 

\subsubsection{Unreal-based Apps}  \label{sec:unreal}
Unreal includes richer information in the \emph{.apk} file than Unity does, including the configurations app developers provide to the development engine regarding their dependence on sensitive data. 
We thus resort to these configuration items to extract the data types the app requests from Unreal. 

Figure~\ref{fig:unreal_static} illustrates the steps conducted in our analysis. 
Unreal stores all app assets~(e.g., models, scenes, game objects, and all plugins from the SDK) in a package named \texttt{main.obb.png}, which is in the zip archive format. 
Unpacking it, we get a  \emph{.pak} file (\ding{172}). 
We then use an Unreal-developed tool UnrealPakTool~\cite{unreal2024parktool} to decode the \emph{.pak} file, resulting in a folder named \texttt{app\_name} (\ding{173}). 
Inside this folder, some apps have a \texttt{Defaultengine.ini} file~(\ding{174}), which stores app requested data in the format of key-value pairs. 
For example, the configuration item \texttt{EnableEyeTracking=True} in Figure~\ref{fig:unreal_static} indicates that the app has activated the eye-tracking feature. 
Other apps have a few \emph{.uplugin} files~(\ding{175}), which contains all the modules called by the app. 
For example, the module \emph{PICOXRMotionTracking} indicates that the app has requested body movement data. 
From these configuration items, we can derive the sensitive data types the app requests. 

\begin{figure}[t]
    \centering
    \includegraphics[width=0.5\textwidth]{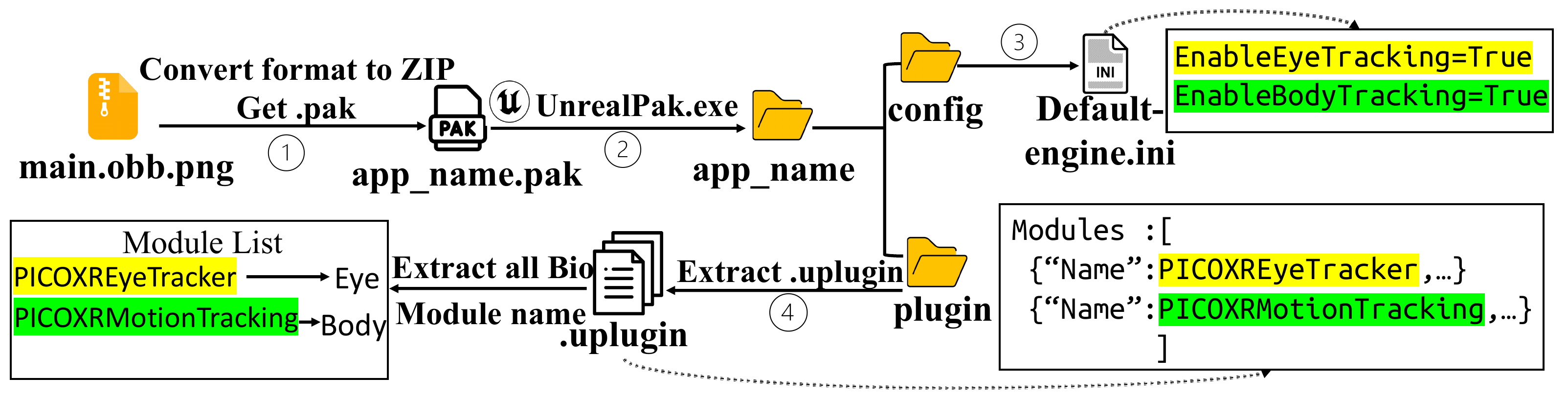}
    \caption{A demonstration of Unreal-based app analysis}
    \label{fig:unreal_static}
\end{figure}

\subsection{Policy Data Types Extraction}
Next, we proceed with extracting data types that the app declares to collect in the privacy policy. 
As several existing studies involve the extraction of entities from privacy policies~\cite{andow2019policylint,xie2022scrutinizing,bui2021automated,harkous2018polisis}, we explore reusing them in our study.
For instance, PolicyLint~\cite{andow2019policylint} generates an ontology and uses a name entity recognition~(NER) model along with specific sentence patterns to accurately extract entities from privacy policies. 
However, in the domain of VR, there is a lack of extensive labeled data and specific sentence patterns needed to train a NER model. 
We thus apply the lightweight method proposed in Skipper~\cite{xie2022scrutinizing}.

Skipper has paragraph and sentence classifiers to obtain sentences related to user data types in the COLLECT component of the privacy policy, such as \emph{``We collect your information include name, email address, state and technical information such as estimated hand size and hand pose data.''} 
After obtaining these sentences, it uses natural language process techniques to parse them to obtain the data types. 
To adapt Skipper for our analysis, we construct a corpus of VR-specific data types from the official documentation of VR platforms and engines. 
Three authors of this paper independently find and label data types into the categories discussed in Section~\ref{sec:spe_data}. 
Then they discuss each collected data type and merge them, leading to the corpus listed in Table~\ref{vr_type}. 
We replace Skipper's original general data dictionary with this VR-specific data corpus, thereby empowering Skipper with the capability to extract VR-specific data from privacy policies. 
To check the precision of Skipper, we randomly select 20 privacy policies that contain a declaration of VR-specific data for manual confirmation. 
Three fails because Skipper mis-labels sentences into \texttt{COLLECT}.

\begin{table}[t]
\caption{VR-specific data types described in privacy policies \label{vr_type}} 
\footnotesize
    \begin{tabular}{l@{\hspace{0.6cm}}r@{\hspace{0.5cm}}}
    \toprule[1pt]
    \textbf{VPD$^\ddagger$} & \textbf{Related Data Types}\\\hline 
    Body & Body Tracking, Motion Capture Data \\ 
    & Physical Interaction Data, User Posture and Movement \\ \hline
    Face & Facial Recognition, Facial Mapping\\
    &Emotion Detection, Facial Geometry Data, Camera\\\hline
    Eye & Eye Tracking, Gaze Detection\\
    &Eye Movement Metrics, Pupil Dilation Data, Iris Scan \\\hline
    Hand & Hand Tracking, Hand Size, Hand Pose Data\\
    &Touch Interaction, Hand Movement Data \\
    \bottomrule[1pt]
    
    \end{tabular}
\begin{flushleft}
    \small
    \textsuperscript{$\ddagger$} VPD: VR-specific Privacy-sensitive Data Types
    
\end{flushleft}

\end{table}


\subsection{Data Access and Privacy Compliance Assessment} 

\begin{table*}[t]
\caption{The status of VR-specific privacy data access and policy compliance}\label{po_result}
\footnotesize
    \begin{tabular}{p{0.05\linewidth}|p{0.15\linewidth}| p{0.06\linewidth} |p{0.03\linewidth}| p{0.13\linewidth}| p{0.13\linewidth}| p{0.13\linewidth}| p{0.13\linewidth}}
    \hline

    \hline
    \textbf{Store} &\textbf{Non-compliance Case} &\textbf{Engine}& \textbf{LF$\dagger$} &\multicolumn{4}{c}{\textbf{Access to VR-specific Privacy-sensitive Data$\S$}} \\ \cline{5-8}
    
    & & & &\textbf{Body}&\textbf{Face}&\textbf{Eye}&\textbf{Hand} \\ \hline
    
    \multirow{4}{*}{Pico} & \multirow{4}{*}{80 (18.3\%)} & Unity & 0  & 17 \hfill (4.9\%) & 8 \hfill (2.3\%) & 6 \hfill (1.7\%) & 22 \hfill (6.4\%)  \\ 
    
     &  &  &   & 1 \hfill (0.3\%) & 0 \hfill (0.0\%) & 2 \hfill(33.3\%) & 3 \hfill (13.6\%)\\ \cline{3-8}
    
                         &  & Unreal & 23  & 7 \hfill (12.7\%) & 7 \hfill (9.0\%) & 0 \hfill (0.0\%) & 21 \hfill (27.0\%)  \\ 
                         
                         &  &  &  & 0 \hfill (0.0\%) & 1 \hfill (14.3\%) &  0 \hfill (0.0\%) & 1 \hfill (4.8\%) \\ \hline
                         
    \multirow{4}{*}{Oculus}&\multirow{4}{*}{64 (17.5\%)}& Unity & 13  &12 \hfill (3.6\%) & 20 \hfill (6.0\%)  & 9 \hfill (2.7\%) & 18 \hfill (5.4\%)\\ 

     &  &  &  & 2 \hfill (16.7\%)& 2 \hfill (10.0\%) & 2 \hfill (22.2\%)& 2 \hfill (11.1\%)\\ \cline{3-8}
    
    & & Unreal & 8  & 2 \hfill (8.7\%)  & 2 \hfill (8.7\%) & 4 \hfill (17.4\%) & 5 \hfill (21.7\%)  \\

     & &  &  & 0 \hfill (0.0\%) & 0 \hfill (0.0\%) & 0 \hfill (0.0\%) & 0 \hfill (0.0\%) \\
     

    \hline

    \hline
    \end{tabular}
\begin{flushleft}
    \small
    \textsuperscript{$\dagger$} LF: The \emph{.apk} lacks resource files.
    \\
    \textsuperscript{$\S$} The upper statistics represents the number of apps that are detected using the corresponding data, and the lower data represents the number of apps that make correct declaration in their privacy policies.
\end{flushleft}
\vspace{-0.25cm}
\end{table*}
We compare the data types extracted from the \emph{.apk} files and the privacy policies. 
Table~\ref{po_result} shows the use of privacy-sensitive data by apps across Pico and Oculus stores.
On the Oculus platform, 331 apps are developed using Unity, 23 with Unreal, and 11 with other game engines.
In Pico, 346 apps are developed with Unity, 78 with Unreal, and 13 with other game engines.
Except for the case of using Unity for development on the Pico platform, there is a lack of resource files in other cases, like \texttt{global\_metadata.dat} in Unity and \texttt{main.obb.png} in Unreal.

We display the frequently occurring VR privacy-sensitive data APIs in Appendix~\ref{app:map}.
Among the 4 VR-specific data types, we find that hand-tracking APIs are among the most frequently used, such as \texttt{GetJointLocations} and \texttt{Pico.Get\_Handness}. These APIs capture hand movements and joint states, enabling gesture control and interactions within the VR environment. Additionally, eye-tracking APIs like \texttt{UPvr\_getEyeTrackingPos} and \texttt{OculusEyeTracker} are widely employed. These APIs track users' gaze directions in real-time, allowing for more natural interactions, such as adjusting perspectives and selecting objects. Similarly, face-tracking APIs, like \texttt{WantFaceTrackingService}, is often used to capture facial expressions, particularly in social VR or virtual character interactions where realistic emotional expression is crucial.


\begin{formal}

\noindent \textbf{Finding \thefinding: \stepcounter{finding}Both Pico and Oculus each have 18.3\% and 17.5\% apps that have misalignment between the declarations in privacy policies and actual data use behaviors.} 
Current VR apps often fail to declare their practices regarding the use of user VR-specific privacy-sensitive data in their privacy policies. Furthermore, even when such declarations are made, we find that most privacy policies use vague terms like ``\emph{biometric data}'', ``\emph{game interaction data}'' or ``\emph{sensor data}'', without specifying the exact types of data involved.

\end{formal}

\section{Discussion}
\subsection{Implications}
The result of our cross-store analysis underscores the existence of significant privacy concerns within the current VR ecosystem. More specifically, the lack of a unified approach to privacy disclosures has led to a fragmented VR ecosystem, with significant variations in user experience and data protection standards. This fragmentation can hinder the overall development of the VR industry, as it complicates efforts to establish universal standards and best practices. Additionally, the VR privacy compliance issues identified in this study result in VR app stores violating data protection regulations, leading to fines or sanctions. In this section, we will further discuss the implications of our research for three stakeholders - VR app developers, store operators, and users.





\paragraph{For store operators} Store operators must enforce a stringent information disclosure process, enabling users to effectively understand the privacy-related information of VR apps.
They should enhance their vetting process to ensure the application is compliant with data protection regulations like GDPR and their privacy policies.  
Platforms should also encourage developers to create specific privacy policies for their apps, rather than relying on generic policies or policies for their other apps, ensuring greater transparency and user trust.

\paragraph{For app developers}
We emphasize the significance of VR app developers prioritizing privacy within the VR ecosystem. In summarizing our findings, we propose the following four guidelines for VR app developers: 
(1) \emph{\textbf{Providing high-quality privacy policies.}}
Developers should create clear and explicit privacy policies for their apps, ensuring that the policies comprehensively outline how data is collected, processed, stored, and shared, including specific details pertaining to VR, such as biometric data and ambient information.
(2) \emph{\textbf{Providing comprehensive app information.}}
This includes offering comprehensive privacy-relevant details on the app's homepage, covering permissions, play style, environmental requirements, and so on. Such information can enhance users' comprehension of the app's data processing practices. Moreover, it is crucial to ensure that the information provided aligns with the actual behavior of the app.
(3) \emph{\textbf{User authorization and option.}}
The app should provide users with a clear authorization mechanism that allows them to selectively consent or decline data collection and sharing. Prioritize respecting user choices and safeguarding their privacy right.
(4) \emph{\textbf{Special attention to children's privacy protection.}}
When the developed app includes children in its user group, the privacy policy should outline a detailed process for handling children's data. 
Otherwise, explicit age restrictions should be provided along with an age verification mechanism to ensure access to the VR app is restricted to users of the appropriate age.

\paragraph{For users}
As users explore the immersive world of VR technology, it is crucial to remain vigilant about personal data security, particularly biometric information. 
Before using any VR app, a thorough review of its privacy policy is imperative. 
This step ensures a clear understanding of the app's practices in collecting, processing, and utilizing personal information. 
Additionally, during the installation of VR apps, it is important to scrutinize the permission requests. 
Users should judiciously consider whether to grant permissions, especially when sensitive information is involved. 

\subsection{Limitations}
To the best of our knowledge, we are the first comprehensive multi-store study of privacy practices of the current VR app ecosystem. 
However, several limitations exist that could be addressed in future work.

\paragraph{Reverse engineering and static analysis}
We rely on static analysis to obtain the behavioral profiles of VR apps. 
It demonstrates the efficiency in finding API invocations to access privacy-sensitive data. 
Additionally, it supports multiple development environments, facilitating the analysis of \emph{.apk} from different manufacturers and engines. 
However, this type of analysis may cause 
false positives, as certain code may not be reachable at runtime. 
This can be enhanced by designing strategies for dynamic analysis in our future work.

\paragraph{Analysis of Android-based apps}
We collect a total of \appnum VR apps from \storenum VR app stores. 
Currently, we only analyze the free-download apps from two Android system app stores (Pico and Oculus) for their behavior due to the limitation in the app availability and toolchains. 

\paragraph{Single privacy policy language analysis}
As shown in Figure~\ref{fig:language}, current VR apps offer privacy policies in many languages. For instance, Pico provides a substantial number of privacy policies in Chinese. However, our analysis currently focuses only on the English versions of these policies. In future work, we plan to include privacy policies in different languages in our analysis.

\vspace{-0.1cm}
\section{Related work}
Previous research has focused on the analysis of VR technology's access to user privacy data~\cite{buck2021privacy,happa2021privacy, lee2022technology,lopez2016method, trimananda2022ovrseen}.
However, few studies in the academic community regarding the privacy compliance of various VR platform apps. To the best of our knowledge, our work represents the largest multi-store study of information and behavior analysis in VR apps.
\paragraph{Privacy concerns in virtual reality}
A series of studies have been dedicated to investigating privacy concerns arising from VR technology~\cite{ginosar2017analytical,lebeck2018towards,happa2021privacy}. 
Adams et al.~\cite{adams2018ethics} present the first work on VR security and privacy perceptions, a mixed-method study involving semi-structured interviews with 20 VR users and developers, a survey of VR privacy policies, and an ethics co-design study with VR developers.
Zhang et al.~\cite{zhang2022peer} propose a multidimensional peer-related privacy concern that focuses on privacy violations from online peers.
Buck et al.~\cite{buck2021privacy} examine the ethical and privacy concerns arising from interactions and the availability of personal data.
Building upon these previous works, we further refine declarative and behavioral concerns to comprehensively assess existing privacy concerns in the current VR ecosystem.

\paragraph{Privacy declaration analysis for applications in different domains}
The emergence of data protection regulations such as GDPR and CCPA, has prompted various application domains to prioritize compliance assessments of privacy declaration information~\cite{yan2024exploring,yan2024investigating,yan2025understanding,yan2025tracking,wang2024corelocker,xie2024your,wan2024analyzing}.
Andow et al.~\cite{andow2019policylint} present PolicyLint, an automated tool for identifying contradictions within Android application privacy policies. PolicyLint employs sentence-level NLP to comprehend the stated declarations regarding data collection and sharing.
In the Virtual Personal Assistant (VPA) domain, Xie et al.~\cite{xie2022scrutinizing} developed Skipper to identify the noncompliance between Alexa skills' behaviors and their declared information in the privacy policy.
Yan et al.~\cite{yan2024on} introduce Quper to perform a comprehensive analysis of the privacy policy quality of VPA applications from four metrics according to data protection regulations.
Drawing from specific terms in VR policies, such as ``eye-tracking'' and ``hand-tracking'' data, we conduct a method for evaluating privacy policies of VR apps based on previous work, ensuring comprehensive protection and compliance for VR user data.

\section{Conclusion}
In this work, we conduct comprehensive large-scale measurements on the privacy practices of the current VR app ecosystem. 
Our study covers \appnum VR apps obtained from \storenum major VR app stores, focusing on their declarative and behavioral privacy profiles. 
We explore the status of VR apps in declaring privacy-related information, evaluate the contents of the declarative documents, and assess the compliance between the app's declarative and behavioral privacy practices.
To our best knowledge, our study represents the inaugural systematic investigation into the \emph{status quo} of privacy practices within VR app ecosystems. 
Our research aims to serve as a crucial alert for VR app developers and a compelling catalyst for VR app store operators to implement robust and effective privacy oversight mechanisms.




\bibliographystyle{IEEEtran}
\bibliography{paper}

{\appendix

\section{Appendix}

\subsection{Other privacy policy analysis}~\label{sec:other_analysis}
\subsubsection{Readability of Privacy Policies}\label{sec:quality}
Privacy policies often lead to \emph{privacy policy fatigue}~\cite{choi2018role} among users due to their complexity and time-consuming nature, necessitating highly readable privacy policies. 
In addition, given that VR apps have a large child user base, it is crucial to develop privacy policies that are easy to understand and suitable for young age groups. 
These policies need to be concise and clear, considering the comprehension ability of child users, to ensure they are directly relevant and easily acceptable to them.

We assess the readability from the word/sentence complexity and the document structure. 
For the former, we employ three widely recognized metrics, i.e., the \emph{automated readability index}~(ARI) that is designed to determine the education grade level needed for document comprehension, \emph{Flesch reading ease score}~(FRES) that evaluates how easy a text is to understand based on sentence length and word complexity, and \emph{Lesbarhetsindex}~(LIX) that assesses the complexity of a text based on the percentage of long words used. 
For the latter, we use seven metrics that effectively capture the essence of a document's structure, including letters per word (LPW), syllables per word (SPW), words per sentence (WPS), sentence count (SC)
, word count (WC), reading time (RT) and speaking time (ST). The results are presented in Figure~\ref{fig:readability}.

\subsubsection{Completeness of Privacy Policies}\label{sec:comp}

\begin{table*}[t]
\small
\caption{A summary of the status of component coverage in privacy policies among different stores\label{tab:completeness}}
    \resizebox{\textwidth}{14mm}{
    \begin{tabular}{l|l|l|l|l|l|l|l|l|l|l}
    \hline

    \hline
     \makecell{\diagbox[linewidth=1pt,width=10.5em]{\textbf{Store}}{\textbf{Component (\%)}}}& \textbf{COLLECT\tnote{1}} & \textbf{SHARE}& \textbf{SECURITY}& \textbf{RIGHT}& \textbf{CHILDREN}& \textbf{REGION}& \textbf{UPDATE}& \textbf{PROVIDER}& \textbf{RETENTION}& \textbf{DATA\_USE}\\\hline 
    \textbf{Oculus} & 96.3 & 76.7& 61.9&88.0 & 76.0&32.7 & 79.6& 79.6& 30.5& 94.6\\ \hline
    \textbf{Viveport} & 96.7 & 77.6& 60.1& 93.2& 79.7& 26.9&70.8&73.1&21.7&92.5\\ \hline
    \textbf{Pico} & 91.8 & 64.3& 47.9& 85.6& 69.2& 28.9& 66.6& 61.3& 19.3& 88.2\\ \hline
    \textbf{Microsoft} & 92.9 & 68.7& 52.1&84.8 & 64.0& 34.1& 58.8& 53.1& 24.2& 87.2\\ \hline
    \textbf{PlayStation} & 90.6 & 68.8& 56.3& 87.5& 71.9& 40.6& 87.5& 90.6& 28.1& 90.6\\ \hline\hline
    \textbf{Total} & 93.7 & 71.2& 55.7& 87.8& 72.2& 32.6& 72.7& 71.5& 24.8& 90.6\\ 
    \hline

    \hline
    \end{tabular}}
\end{table*}



The completeness of a privacy policy pertains to how thoroughly its disclosed information meets the requirements set by regulations, store operators, and users. 
We take as reference GDPR Article 13~\cite{gdpr}, which outlines the necessary components of a privacy policy. 
Moreover, recent studies~\cite{andow2019policylint,completeness2015a, yan2024on,liu2021have} have summarized the types of content derived from data regulations or user studies. 

We utilize QuPer~\cite{yan2024on}, a tool that automatically labels paragraphs within the privacy policies into components. 
It supports eleven components, including \texttt{COLLECT}, \texttt{SHARE}, \texttt{SECURITY}, \texttt{RIGHT}, \texttt{CHILDREN}, \texttt{REGION}, \texttt{UPDATE}, \texttt{PROVIDER}, \texttt{RETENTION}, \texttt{DATA\_USE} and \texttt{COOKIE}, which are shown in Table~\ref{tab:ppcategory}.
We remove the \texttt{COOKIE} component, as VR apps are not relevant to browser cookies. 
Quper is claimed to achieve an F1-score of 82\%. 
We also randomly select fifty privacy policies and manually confirm the identified components by Quper. 
In Table~\ref{tab:completeness}, we summarize the coverage status of the privacy policies among the five app stores. 


\begin{table*}[t]
\caption{The mapping between the privacy-sensitive data~(Section~\ref{sec:spe_data}) and APIs}\label{pico_oculus_api}
\small
    \begin{tabular}{p{0.04\linewidth}p{0.04\linewidth} p{0.04\linewidth}p{0.78\linewidth}}
    \toprule[1pt]

    \textbf{Store} & \textbf{Engine} & \textbf{VPD$^\ddagger$} & \textbf{APIs/Classes}\\\hline 
     Pico     &    Unity   & Body  & \texttt{GetBodyTrackingPose, BodyTrackerRole, BodyTrackerResult, BodyTrackerTransform} \\ \cline{3-4}
          &       & Face  & \texttt{WantFaceTrackingService, GetFaceTrackingSupported, StartFaceTracking} \\ \cline{3-4}
          &       & Eye  & \texttt{UPvr\_getEyeTrackingPos, UPvr\_getEyeTrackingData, UPvr\_getEyeTrackingGazeRay} \\ \cline{3-4}
          &       & Hand  & \texttt{GetHandScale, GetJointLocations, GetSettingState} \\ \cline{2-4}

          &   Unreal     & Body  & \texttt{PXR.Get\_Body\_Tracking\_Pose, PXR.Set\_Swift\_Mode, PICOXRMotionTracking}  \\ \cline{3-4}
          &        & Face  & \texttt{Pico.Get\_Face\_Tracking\_State, Pico.Start\_Face\_Tracking, EnableFaceTracking}  \\ \cline{3-4}
          &        & Eye  & \texttt{Pico.Get\_Eye\_Tracking\_Gaze\_Ray, Pico.Set\_Boundary\_Visible, PICOXRHMD,OpenXREyeTracker}  \\ \cline{3-4}
          &        & Hand  & \texttt{Pico.Get\_Handness, PicoMobileController, PICOXRHMD, OpenXRHandTracking}  \\ \hline

    Oculus       &    Unity   & Body  & \texttt{OVRBody, OVRBone, OVRCustomSkeleton}\\ \cline{3-4}
           &       & Face  & \texttt{OVRCustomFace, OVRCustomFaceExtensions, OVRFace}\\ \cline{3-4}
           &       & Eye  & \texttt{OVREyeGaze}\\ \cline{3-4}
           &       & Hand  & \texttt{OVRHand}\\ \cline{2-4}

           & Unreal       & Body  & \texttt{OpenXRHMD, OpenXREditor, OpenXR} \\ \cline{3-4}
           &        & Face  & \texttt{FacialAnimation}\\ \cline{3-4}
           &        & Eye  & \texttt{OculusEyeTracker} \\ \cline{3-4} 
           &        & Hand  & \texttt{GetHandJointTransform, GetHandJointTransform}\\ 
           
    \bottomrule[1pt]
    \end{tabular}
\begin{flushleft}
    \small
    \textsuperscript{$\ddagger$} VPD: VR-specific Privacy-sensitive Data Types
\end{flushleft}
\vspace{-0.2cm}
\end{table*}


\begin{table*}[t] 
\centering

\setlength{\tabcolsep}{1.5pt}

\caption{Main components in VR privacy policy}
\label{tab:ppcategory} 
\small
  \begin{tabular}{p{0.02\linewidth} | p{0.095\linewidth}| p{0.245\linewidth}| p{0.6\linewidth} }
 \hline

 \hline
 & \textbf{Component} & \textbf{Description} & \textbf{Example}\\
\hline
 1 & COLLECT & What types of data does the app collect from the user & We collect your information include name, email address, state and technical information such as estimated hand size and hand pose data.\\
 \hline
 2 & SHARE & How the app shares user information & We may share your personal data with third parties. \\  
 \hline
 3 & SECURITY & How the app protects user information & These security measures encompass password-protected directories and databases to secure your information, along with SSL (Secure Sockets Layer) technology to guarantee full encryption of your data. \\ [1ex]
 \hline
 4 & RIGHT & Users' rights to their own data & You have the right to stop the advertising messages that we send to you at any time. \\  
 \hline
 5 & CHILDREN & Privacy Policy for Child Protection Mechanisms & We do not collect any information from anyone under 13 years of age. \\ 
 \hline
 6 & REGION & Protection Mechanisms for Some Special Regions &
 If you are a resident of the state of California, we will abide by the regulations of CalOPPA when handling your information.\\ 
 \hline
 7 & UPDATE & Information about privacy policy updates & Be aware that this Privacy Policy is subject to periodic updates. For the most current version in effect, please consult our website. \\ 
 \hline
 8 & PROVIDER & Contact information of the privacy policy provider & If you have questions of your personal data, please send email to us at: xxx@.com.\\ 
 \hline
 9 & RETENTION & How long will the app keep user data & We will retain your data for 12 months.\\ [1ex]
 \hline
 10 & DATA\_USE & How the app will use user data & The data we gather is utilized to enhance our website, aiming to provide you with improved service.\\
 \hline

 \hline
\end{tabular} 
\vspace{-0.1cm}
\end{table*} 
\begin{figure*}[t]
    \centering
    \includegraphics[width=0.9\textwidth]{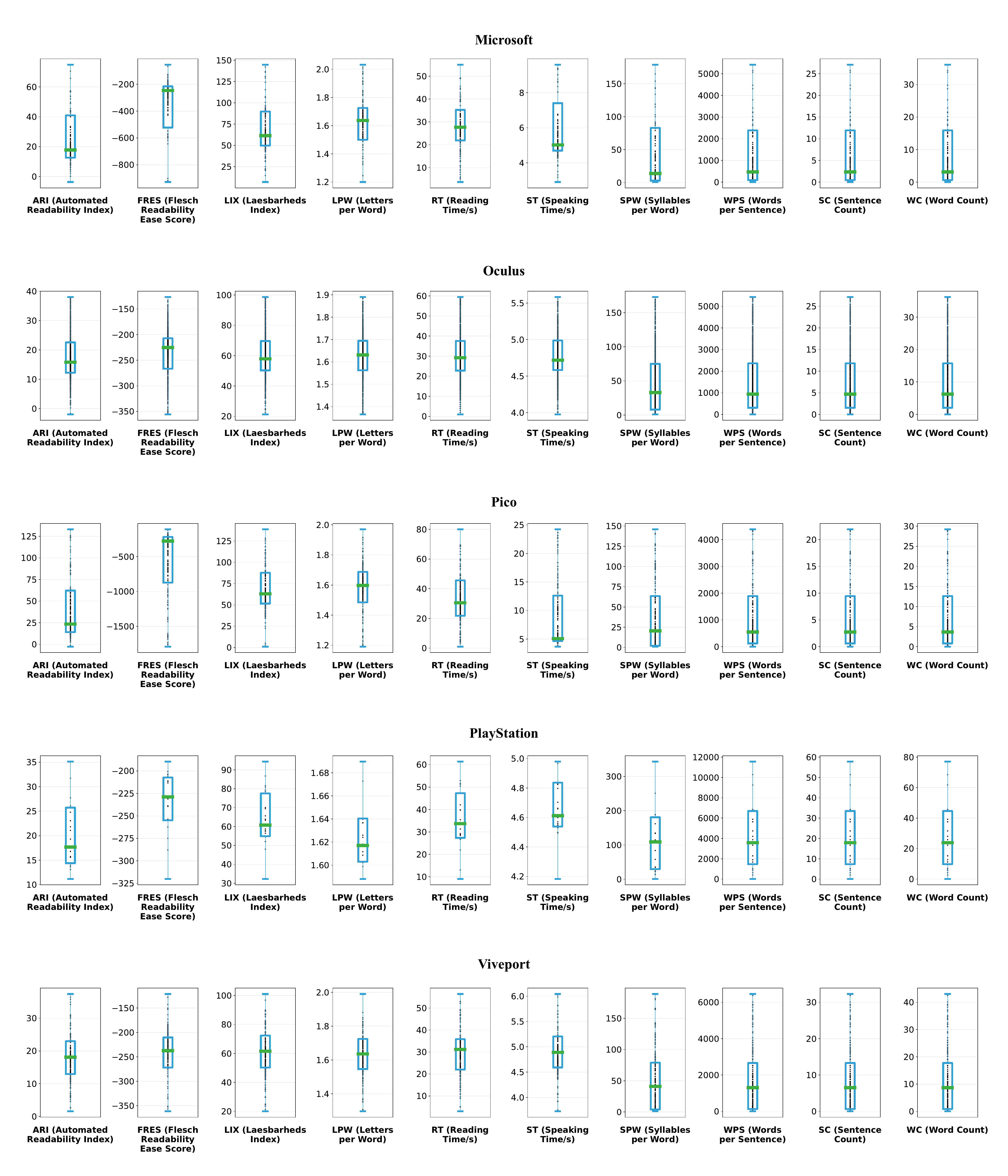}
    \caption{Result of readability metrics in 5 VR app stores privacy policy}
    \label{fig:readability}
\end{figure*}

\subsection{Mapping between the privacy-sensitive data and APIs} \label{app:map}
Table~\ref{pico_oculus_api} shows the summary of APIs used to access VR-specific privacy-sensitive data from the official SDK documentation of Pico and Oculus.

 }

 
%


 




\vfill

\end{document}